\documentclass[conference]{IEEEtran}
\IEEEoverridecommandlockouts

\usepackage{cite}
\usepackage{amsmath,amssymb,amsfonts}
\usepackage{algorithmic}
\usepackage{graphicx}
\usepackage{textcomp}
\usepackage{xcolor}
\usepackage{booktabs}
\usepackage{multicol}
\usepackage{multirow}
\usepackage{amsfonts}
\usepackage{colortbl}
\usepackage{url}
\usepackage{hyperref}
\def\BibTeX{{\rm B\kern-.05em{\sc i\kern-.025em b}\kern-.08em
    T\kern-.1667em\lower.7ex\hbox{E}\kern-.125emX}}
\begin{document}

\title{Robust Training of Singing Voice Synthesis Using Prior and Posterior Uncertainty
}

\author{
    \IEEEauthorblockN{Yiwen Zhao\IEEEauthorrefmark{2}, Jiatong Shi\IEEEauthorrefmark{2}, Yuxun Tang\IEEEauthorrefmark{1}, William Chen\IEEEauthorrefmark{2}, Shinji Watanabe\IEEEauthorrefmark{2}}
    \IEEEauthorblockA{\IEEEauthorrefmark{2}Carnegie Mellon University, \IEEEauthorrefmark{1}Renmin University of China}
}


\maketitle

\begin{abstract}
Singing voice synthesis (SVS) has seen remarkable advancements in recent years. However, compared to speech and general audio data, publicly available singing datasets remain limited. In practice, this data scarcity often leads to performance degradation in long-tail scenarios, such as imbalanced pitch distributions or rare singing styles. To mitigate these challenges, we propose uncertainty-based optimization to improve the training process of end-to-end SVS models. First, we introduce differentiable data augmentation in the adversarial training, which operates in a sample-wise manner to increase the prior uncertainty. Second, we incorporate a frame-level uncertainty prediction module that estimates the posterior uncertainty, enabling the model to allocate more learning capacity to low-confidence segments. Empirical results on the Opencpop and Ofuton-P, across Chinese and Japanese, demonstrate that our approach improves performance in various perspectives.

\end{abstract}


\section{Introduction}
\label{sec:introduction}
Singing voice synthesis (SVS) aims to generate natural and expressive singing vocals conditioned on input lyrics, pitch, and duration~\cite{cook1996singing,hono2021sinsy,kenmochi2007vocaloid}. The task has gained increasing popularity due to growing demands for digital human applications~\cite{yu2017talking,kearney2016design} and the rapid advancement of deep generative modeling techniques~\cite{NIPS2014_f033ed80,kingma2013auto,song2020denoising}. Compared to text-to-speech (TTS)~\cite{tomoki2022espnettts,cosyvoice2024du} or general music generation~\cite{bai2024seedmusicunifiedframeworkhigh,zhang2025inspiremusicintegratingsuperresolution}, SVS presents unique challenges: it requires accurate alignment with linguistic content while also capturing fine-grained prosody, making it particularly complex for neural models to learn effectively. Furthermore, the development of SVS is hindered
by limited data availability: SVS models tend to underperform in long-tail scenarios, such as high-pitch phrases or underrepresented
singing styles. In contrast to speech or music
datasets, publicly available singing corpora remain scarce. Licensing songs from professional artists is often difficult, and annotating MIDI scores requires expert-level effort. 

Previous work has explored data augmentation and corpus expansion to address these data scarcity issues~\cite{guo2022singaug, shi2024singing, ghosh2024synthio, wang2024prompt, yu2024visinger2+,ren2020deepsinger,shi2021sequence,saino2006hmm,wang2022opencpop,ogawa2021tohoku}. The motivation is to provide more data variation, enabling the model to leverage samples with higher uncertainty (i.e., diverse pitch range or rare musical style) and update its gradient more efficiently. However, these approaches typically aim to avoid introducing excessive noise or artifacts to ensure that the augmented data remains close to the original distribution. While this helps preserve generation quality, it may also prevent generalizability to more diverse or challenging singing patterns. The data expansion and conservative augmentations introduce variations in a fixed scale or small range, and constrain the optimization to modifying the data. However, equipping the model with a better awareness of its own uncertainty can enhance robustness even under limited data conditions, by guiding learning toward more challenging regions, which is a potential that remains underexplored.


In this work, we introduce two uncertainty-based optimization methods to improve the training process: (1) a \textbf{differentiable augmentation module} in adversarial training to increase the \textit{prior uncertainty}, and (2) a \textbf{uncertainty prediction module} that estimates the model's \textit{posterior uncertainty} using the frame-level latents. The \textit{prior} means formulate without introducing the true value, and the \textit{posterior} indicates with the true value, which we have a detailed explanation on Sec.~\ref{sec:formulation}. Together, these techniques enrich the training process and help the model better handle long-tail cases. 
As a result, we observe consistent performance improvements over the baseline. We provide a detailed analysis in Sec.~\ref{sec:quantitativea} to better understand how the impact of each component varies depending on the characteristics of the corpus. Additionally, these uncertainty formulations can be seamlessly combined with existing SVS models, which commonly employ GAN-based vocoders~\cite{kong2020hifi, yamamoto2020parallel} or two-stage joint adversarial training~\cite{asystematic}, and with reconstruction-based objectives.

In summary, our contributions include:



\begin{itemize}
    \item We are the first to apply differentiable data augmentation to SVS, enabling stronger input perturbation while preserving the target distribution unchanged for generation.
    \item We propose a posterior uncertainty modulation for sample-level audio interval, demonstrating its effectiveness in enhancing robustness.
    \item We integrate these uncertainty-based techniques into an SOTA end-to-end SVS model, achieving consistent improvements across multiple metrics.
\end{itemize}
We provide singing samples from different systems on a website: \href{https://tsukasane.github.io/SingingUncertainty/}{https://tsukasane.github.io/SingingUncertainty/}

\section{Related Work}

\subsection{Singing Voice Synthesis}
\label{sec:relatedwork_svs}
Early deep learning-based singing voice synthesis (SVS) systems typically adopt a two-stage pipeline~\cite{blaauw2020sequence,lu2020xiaoicesing,chen2020hifisinger,liu2022diffsinger}: the model first predicts intermediate acoustic features, which are then converted into waveforms using a vocoder. While this approach has proven effective, it often suffers from information loss between stages and limited expressiveness in the final output. More recently, end-to-end SVS models such as VISinger~\cite{zhang2022visinger} and VISinger2~\cite{zhang2022visinger2} have gained increasing attention. These models leverage a combination of conditional variational autoencoders~(CVAEs)\cite{kingma2013auto} and generative adversarial networks~(GANs)\cite{NIPS2014_f033ed80} to directly generate waveforms from music score inputs. This fully end-to-end architecture simplifies the training pipeline and improves synthesis fidelity. VISinger2+~\cite{yu2024visinger2+} concatenates discrete representations obtained from pre-trained SSL models~\cite{hsu2021hubert,li2023mert} to the input mel-spectrogram, further improving the performance in musical details and singer similarity in a multi-singer corpus~\cite{shi2024singing}. Despite these advancements, SVS remains a challenging task due to several factors, (1) \textbf{fluctuating pitches} in techniques such as vibrato. (2) \textbf{sensitive timing} alignment when dealing with expressive timing or tempo variations. (3) \textbf{expressiveness} for underrepresented singing styles, emotions, or vocal techniques. Our uncertainty-based training method helps address these issues, improving the musicality and expressiveness of the generated audio while being more data-efficient and robust to long-tail cases.


\begin{figure*}[t]
\centering
\includegraphics[width=17.5cm]{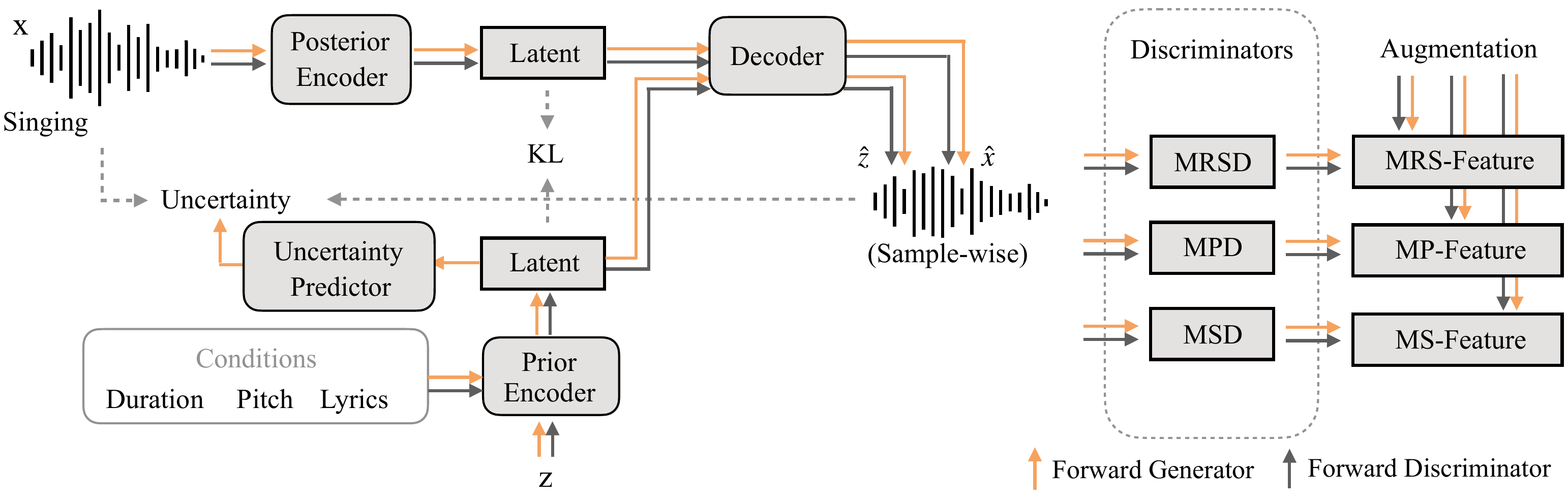}
\vspace{-3mm}
\caption{The pipeline of VISinger2 with our proposed uncertainty predictor and differentiable augmentation. The uncertainty predictor is trained with a forward generator pass, and the differentiable augmentation is added to spectrogram-level features in adversarial training.}
\label{fig_pipeline}
\vspace{-3mm}
\end{figure*}

\subsection{Data Augmentation}
Previous works have explored several augmentation and data expansion strategies to mitigate the data scarcity issue. Singaug~\cite{guo2022singaug} introduces pitch augmentation by simultaneously shifting the music score and the waveform by a certain number of semitones. They also propose feature space mixup and add a corresponding loss based on the mixed acoustic target. Other works~\cite{choi2022melody, wang2024prompt} leveraged out-of-domain speech data or low-quality singing recordings with background noise as auxiliary training resources. There are also approaches that uses commercial SVS systems, such as ACE Studio, to synthesize stylized singing and use in training~\cite{kenmochi2007vocaloid,shi2024singing}.



While these methods can ameliorate data scarcity, the methods target generation tasks typically prioritize maintaining the original training distribution. In speech understanding tasks, however, previous works have shown that simple augmentation to the speech data distribution could significantly improve the speech recognition performance~\cite{park2019specaugment}. 
To apply this similar concept to the synthesis domain, we propose to use a differentiable singing augmentation method that introduces more substantial perturbations to the end-to-end SVS training process. The model is provided with diverse real-fake feature pairings, increasing the data variation and therefore prior uncertainty. In addition, we leverage an uncertainty predictor to improve the model's awareness of posterior uncertainty and improve its robustness.




\subsection{Uncertainty Prediction}

Uncertainty prediction is widely used across machine learning domains to enhance model performance in various tasks~\cite{shi2022investigation, kendall2017uncertainties, kuleshov2018accurate,gawlikowski2023survey,hu2024cg}. Uncertainty can generally be categorized into data uncertainty (aleatoric) and model uncertainty (epistemic)~\cite{gawlikowski2023survey}. Recent works on NeRF and Gaussian Splatting for 3D reconstruction~\cite{hu2024cg,huang2024gaussianmarker,pan2022activenerf,shen2022conditional} frequently incorporate uncertainty estimation as a core component. In these contexts, covariance matrices are commonly used to capture data uncertainty, representing the sensitivity of rendering results to input perturbations or spatial variance. On the other hand, Jacobian-based methods are often employed to measure model uncertainty, quantifying the network's confidence in its predictions. However, applying Jacobian-based uncertainty estimation to audio signals is challenging, as the outputs are long temporal sequences, leading to unstable or excessive measurements. Previous work~\cite{shi2022investigation} on target-speaker recognition introduces a neural uncertainty estimator that leverages confidence scores from the speaker extraction module to enhance recognition performance, demonstrating the potential of using auxiliary predictors to modulate uncertainty. Similarly, prior studies in classification and regression~\cite{kendall2017uncertainties, kuleshov2018accurate} have used error-based proxies to model posterior uncertainty more effectively. Motivated by these approaches, we propose to modulate both prior uncertainty and posterior uncertainty in our SVS framework, aiming to enable more efficient and robust training.


\section{Formulation}
\label{sec:formulation}

Conditioned on the music score $M:= (M^{\text{ph}}, M^{\text{pi}}, M^{\text{du}})$, SVS targets to generate a human singing phrase $\mathbf{\hat{x}} \in \mathbb{R}^{T}$ that align with $M$, where $T$ is the length of the waveform, $M^{\text{ph}}\in 
 \mathbb{R}^{T'}$, $M^{\text{pi}}\in 
 \mathbb{R}^{T'}$, $M^{\text{du}}\in 
 \mathbb{R}^{T'}$ represents information about phoneme, pitch, and duration over sequences of the same length $T'$. Training data $\mathbf{x} \in \mathbb{R}^{T}$ is the singing phrase in the source distribution. We formulate the augmented SVS as a multi-stage training with a variety of strategies. In this section, we first introduce the baseline framework, then elaborate on differentiable augmentation, and lastly detail the architecture and mathematical definition of the uncertainty predictor.

\subsection{VISinger2 Framework}

VISinger2~\cite{zhang2022visinger2} is an end-to-end SVS model built upon CVAE and GAN architecture as shown in Fig.~\ref{fig_pipeline}. The input waveform $\mathbf{x}$ is first converted to a mel-spectrogram. The posterior encoder\footnote{Notably, the prior (w/o true value) and posterior (w/ true value) in uncertainty are defined differently from the prior  (w/o known distribution) and posterior (w/ known distribution) used in CVAE explanation.}, as shown in the upper-left of Fig.~\ref{fig_pipeline}, takes the mel-spectrogram and extracts the latent representation $\mathbf{l_x}$.
\begin{equation}
    \mathbf{l_x}=\text{Enc}_{\text{post}}(\text{Mel}(\mathbf{x}))\sim q_{\text{post}}(\mathbf{l_x}|\mathbf{x}),
    \label{eq:posterior_enc}
\end{equation}
where $q_{\text{post}}$ denotes the posterior distribution. The decoder is trained to reconstruct the waveform.
\begin{equation}
    \mathbf{\hat{x}}=\text{Dec}(\mathbf{l_x})\sim p(\mathbf{x}|\mathbf{l_x}).
    \label{eq:posteriorDec}
\end{equation}
The prior encoder, as shown at the bottom-left of Fig.~\ref{fig_pipeline}, takes the music score $M$ and a prior $\mathbf{z}\sim \mathcal{N}(\mathbf{0},\mathbf{I})$ sampled from a standard Gaussian (or flow model modified distribution) to generate a prior latent $\mathbf{l_z}$.
\begin{equation}
    \mathbf{l_z}=\text{Enc}_{\text{pri}}(\mathbf{z}|M)\sim q_{\text{pri}}(\mathbf{l_z}|\mathbf{z},M).
    \label{eq:priorEnc}
\end{equation}
Then, the shared decoder also reconstructs the waveform using the prior latent as
\begin{equation}
    \mathbf{\hat{z}}=\text{Dec}(\mathbf{l_z}). 
    \label{eq:zhatlz}
\end{equation}
The prior latent distribution in Eq.~\eqref{eq:priorEnc} is drawn to the posterior latent distribution in Eq.~\eqref{eq:posterior_enc} using the KL loss. Then, the decoder of the CVAE serves as a generator $G$, where several discriminators $D$ take the posterior sample $\mathbf{\hat{x}}$ and prior sample $\mathbf{\hat{z}}$ for adversarial training. 


\subsection{Differentiable Augmentation}
\label{sec_diffaug}

\noindent\textbf{Adversarial Training}

\label{sec_adversarialtraining}
GAN-based end-to-end models are highly susceptible to mode collapse when training on limited amounts of data. The discriminator tends to memorize the training samples instead of learning meaningful features to distinguish the real and generated samples. Differentiable augmentation has been utilized in image generation~\cite{zhao2020differentiable} to enhance data diversity in StyleGAN~\cite{karras2019style} training. We also adopt this technique for SVS to facilitate continuous updates to the generator. In the baseline model VISinger2~\cite{zhang2022visinger2}, three types of discriminators are used, namely Multi-Resolution Spectrogram Discriminator (MRSD), Multi-Period Discriminator (MPD), and Multi-Scale Discriminator (MSD) as shown on the right side of Fig.~\ref{fig_pipeline}. They capture features in the frequency and time domains across multiple scales. To simplify the notation, we use $D$ to collectively represent these three discriminators. The adversarial loss in this case is formulated as:
\begin{align}
    \mathcal{L}_D = &\mathbb{E}_{\mathbf{l_x} \sim q_{\text{post}}(\mathbf{l_x})} \left[ (1-D(\mathbf{l_x})) \right] + \nonumber\\
    &\mathbb{E}_{\mathbf{l_z} \sim q_{\text{pri}}(\mathbf{l_z}|\mathbf{z},M)} \left[ (D(G(\mathbf{l_z})) \right],
    \label{eq:discriminator}
\end{align}
\begin{align}
    \mathcal{L}_G = &\mathbb{E}_{\mathbf{l_z} \sim q_{\text{pri}}(\mathbf{l_z}|\mathbf{z},M)} \left[(-D(G(\mathbf{l_z}))) \right],
    \label{eq:generator}
\end{align}
where the discriminators $D$ are trained to distinguish between real and synthesized samples generated from different latents, and the generator $G$ (also as the decoder of CVAE) aims to deceive the discriminator. 

\noindent\textbf{Augmentation Methods}

Previous augmentations in automatic speech recognition~(ASR) and speech enhancement (SE) include masking, adding noise, and warping~\cite{park2019specaugment, ghosh2024synthio}. These aggressive augmentations can be adapted in adversarial training, without shifting the original data distribution of $p(\mathbf{\hat{x}})$~\cite{zhao2020differentiable}. We use $A$ to denote these augmentation operations. We introduce the formulation of differentiable masking and adding noise, then show the result of using them individually or in combination in \ref{sec_ablation}. In practice, these techniques can be selected based on task- or dataset-specific characteristics.

We denote the encoders of the three discriminators as $\text{Enc}_D$, then the posterior features $\mathbf{\mathbf{f_{\hat{x}}}}$ and prior features $\mathbf{\mathbf{f_{\hat{z}}}}$ extracted by $\text{Enc}_D$ are:
\begin{equation}
    \mathbf{f_{\hat{x}}} = \text{Enc}_D(\mathbf{\hat{x}}),\quad
    \mathbf{\mathbf{f_{\hat{z}}}} = \text{Enc}_D(\mathbf{\hat{z}}),
\end{equation}
which corresponds to the right part of Fig.~\ref{fig_pipeline}. As observed in~\cite{zhao2020differentiable}, applying augmentations directly to posterior features $\mathbf{f_{\hat{x}}}$ can lead the model to learn the distribution of $A({\mathbf{\hat{x}}})$ rather than $\mathbf{\hat{x}}$, introducing a distributional shift in generated samples. On the other hand, if augmentation is applied solely during the discriminator's forward pass, the discriminator may only learn to distinguish between $A(\mathbf{f_{\hat{x}}})$ and $A(\mathbf{f_{\hat{z}}})$, rather than focusing on the $\mathbf{f_{\hat{x}}}$ and $\mathbf{f_{\hat{z}}}$ distinction. Therefore, we apply augmentation to both $\mathbf{f_{\hat{x}}}$ and $\mathbf{f_{\hat{z}}}$ during the generator and discriminator forward pass as $A(\mathbf{f_{\hat{x}}})$ and $A(\mathbf{f_{\hat{z}}})$. The training flow is shown as the orange and black arrows in Fig.~\ref{fig_pipeline}. To simplify the notations, we denote $\mathbf{f_{\hat{x}}}$ and $\mathbf{f_{\hat{z}}}$ all as $\mathbf{f_D}$ start from here. We explain the operations $A(\mathbf{f_D})$ as follows.
 
Rather than corrupting the entire feature segment, we only target a randomly selected partition. For the time dimension, we first compute the augmented target length based on a predefined ratio $r$, then select a random start index $t_0 \in [1, T_f - \Delta t]$
\begin{equation}
\Delta t = \lfloor T_f \cdot r \rfloor,
\end{equation}
where $T_f$ is the temporal length of the feature. Adding a mask along the time dimension to the sample interval can be formulated as:
\begin{equation}
B_t = 
\begin{cases}
v, & \text{if } t \in [t_0,...,t_0+\Delta t), \\
1, & \text{otherwise}.
\end{cases}
\end{equation}
\begin{equation}
    A(\mathbf{f_D}) = \text{Mask}_{\text{temp}}(\mathbf{f_D},\{B_t\}_t).
\end{equation}
where $B_t$ is a binary mask $B_t \in \{v, 1\}$, with fixed value $v \in [0, 1)$. $B_t$ corrupts the randomly selected interval along the time dimension to obtain $A(\mathbf{f_D})$ by the $\text{Mask}_{\text{temp}}(\cdot)$ operation, which is the masked feature. This operation is differentiable to $\mathbf{f_D}$ and also to the input $\mathbf{x}$. 

Similarly, we define the binary mask at frequency bin $\xi$ as $B_\xi \in \{v, 1\}$. The target frequency interval is also randomly selected. The masked STFT representation is obtained by element-wise multiplication of the original feature
\begin{equation}
B_\xi = 
\begin{cases}
v_\xi, & \text{if } \xi \in [\xi_0, ...,\xi_0 + \Delta \xi), \\
1, & \text{otherwise}.
\end{cases}    
\end{equation}
\begin{equation}
    A(\mathbf{f_D}) = \text{Mask}_{\text{freq}}(\mathbf{f_D},\{B_\xi\}_{\xi}).
\end{equation}
The $\text{Mask}_{\text{freq}}(\cdot)$ operation corrupts a randomly selected band of frequencies in the signal and is differentiable with respect to $\mathbf{x}$. In practice, when we choose masking for differentiable augmentation, we use a frequency dimension mask on the output feature of the MRSD encoder, and a time dimension mask on outputs of the MPD and MSD encoders.

Adding noise is another way to inject controlled perturbations into the waveform input during training. It enables the model to learn from a wider range of acoustic variations. Given an input feature $\mathbf{f_D}$, adding noise $N$ to the feature can be formulated as:
\begin{equation}
    A(\mathbf{f_D}) = \text{Noise}(\mathbf{f_D},N).
\end{equation}
In the randomly selected interval, $N$ is the noise sampled from a standard Gaussian multiplied by a noise scaling factor $\alpha$. The $\text{Noise}(\cdot)$ operation adds this noise to $\mathbf{f_D}$.

Finally, the noisy segment is padded and stitched back into the original features to form the final augmented batch. This local perturbation adds noise variations while retaining most of the original feature content, improving the model’s ability to generalize to unseen scenarios. The masking and adding noise strategy can also be used simultaneously, which we will discuss in Sec.~\ref{sec_ablation}.

As shown in Fig.~\ref{fig_pipeline}, the augmentation $A(\mathbf{f_D})$ introduces diverse real-fake feature pairs and optimizes the data uncertainty, allowing the generator to efficiently update. We denote this as a prior uncertainty, which is distributed according to $p(\mathbf{\hat{x}})$ and $p(\mathbf{\hat{z}})$. Incorporating it with adversarial training keeps consistency in the learning target and enhances the model’s ability to generalize.

\subsection{Uncertainty Prediction}
\label{sec_uncertainty_predictor}

The posterior uncertainty refers to the error proxies, which show the confidence of the model prediction. Rather than being merely undesirable, it can be explicitly modeled to improve prediction reliability~\cite{gawlikowski2023survey,kendall2017uncertainties}. Given variable-length frame-level latent representations, we introduce a convolution-based uncertainty predictor to estimate the uncertainty level across the generated sample-wise outputs.

As defined in Eq.~\eqref{eq:priorEnc}, the frame-wise latent $\mathbf{l_z}$ is passed to the generator (or the decoder of CVAE) and further converted to the prior sample as in Eq.~\eqref{eq:zhatlz}. We define a predictor $g(\mathbf{l_z})$ that takes the latent as input and predicts the uncertainty $\mathbf{u}$. 
\begin{equation}
    \mathbf{u} = g(\mathbf{l_z}).
\end{equation}
The frame-level latent is first linearly interpolated to a fixed interval length, then passes through two 1D convolutional layers for feature transformation, and finally uses a fully connected layer to predict a single scalar value $u_t$ for each time step, referring to uncertainty.

This additional predictor requires a pre-trained model that is already capable of generating reasonable output, so that error prediction can be meaningful. In training, this uncertainty is supervised by the ground truth L2 distance between the input $\mathbf{x}$ and the reconstructed sample $\mathbf{\hat{z}}$. 
\begin{equation}
\mathbf{d} = || \mathbf{x} - \mathbf{\hat{z}} ||^2,
\end{equation}
\begin{equation}
\mathcal{L}_u = \frac{1}{n} \sum_{t=1}^{n} (d_t - u_t)^2,
\end{equation}
$n$ is the total number of samples in an interval. In this manner, the uncertainty predictor is trained to quantify the model's confidence with the true value $\mathbf{x}$ provided, thus it is posterior. 


\begin{table*}[t]
\centering
\caption{Comparison of different training strategies on the Opencpop corpus in 200 epochs. "B" indicates baseline, "U" indicates uncertainty prediction, and "D" represents differentiable augmentation. "\&" means resume and "+" means simultaneously use. MOS values are presented with 95\% confidence intervals.}
\vspace{-5pt}

\footnotesize
\setlength{\tabcolsep}{2.1 mm}
\begin{tabular}{lccccccccc}
\toprule
\multirow{2}{*}{Strategies} & \multicolumn{9}{c}{Opencpop} \\ \cmidrule(r){2-10} & Log\_F0\_RMSE$\downarrow$ & MCD$\downarrow$ & Semitone$\uparrow$ & VUV$\downarrow$ & Sheet-SSQA$\uparrow$ & $\text{MOS}_{L}\uparrow$ & $\text{MOS}_{M}\uparrow$ & $\text{MOS}\uparrow$ & \\ \midrule \midrule
$\textbf{B}$   & 0.174 & 7.876 & 62.78\% & 7.79\% & 4.30 & 3.65 ±0.07 & 3.37 ±0.07 & 3.46 ±0.06\\
$\textbf{B+D}$  & \textbf{0.159} & \textcolor{blue}{7.670} & \textcolor{blue}{63.64\%} & \textcolor{blue}{7.32\%} & \textbf{4.33} & \textcolor{blue}{3.73 ±0.07} & \textcolor{blue}{3.46 ±0.07} & 3.53 ±0.06 \\
$\textbf{B\&U}$  & 0.173 & 7.776 & 63.62\% & 7.37\% & \textcolor{blue}{4.31} & 3.72 ±0.07 & 3.45 ±0.07 & \textcolor{blue}{3.54 ±0.06} \\
$\textbf{B\&U\&(U+D})$  & \textcolor{blue}{0.168} & \textbf{7.659} & \textbf{64.56\%} & \textbf{6.96\%} & 4.30 & \textbf{3.90 ±0.07} & \textbf{3.64 ±0.07} & \textbf{3.67 ±0.06}\\
\midrule
\bottomrule
\end{tabular}
\label{tab_compare}
\vspace{-2mm}
\end{table*}



\section{Experiment}
\label{sec:exp_training_strategies}
\subsection{Training Strategies}

\noindent\textbf{Baseline Architecture}

Our training process consists of multiple stages, with the following annotation conventions: + denotes the simultaneous application of components, while \& indicates a resumed training stage. \textbf{B} represents the baseline model, whereas \textbf{D} and \textbf{U} correspond to the differentiable augmentation and uncertainty prediction, respectively, both of which are designed to enhance training process. The main experiments are conducted on the Opencpop corpus over 200 training epochs.

We use VISinger2~\cite{zhang2022visinger2} implemented in Muskits-ESPnet~\cite{wu2024muskits, shi2022muskits} as our baseline. The prior encoder in Eq.~\eqref{eq:priorEnc} conditioned on music score comprises 6 blocks, each featuring a 2-head relative self-attention mechanism and a 1D-convolutional feed-forward network (FFN). The decoder component in Eq.~\eqref{eq:posteriorDec} upsamples the features. Weight normalization is applied in both the decoder and the posterior encoder. We do not use the flow network to transform the prior distribution so that it is a standard Gaussian as in Eq.~\eqref{eq:priorEnc}. The discriminator part incorporates multi-period, multi-scale, and multi-frequency components as detailed in Sec.~\ref{sec_adversarialtraining}. The generator and discriminator, as in Eq.~\eqref{eq:generator} and Eq.~\eqref{eq:discriminator} are trained using the AdamW optimizer~\cite{AdamW2019Loshchilov} with an initial learning rate of 2.0e-4 and an exponential learning rate scheduler with a gamma of 0.998. More detailed configuration refers to ESPnet~\cite{watanabe2018espnet} config\footnote{\url{https://github.com/espnet/espnet/blob/master/egs2/opencpop/svs1/conf/tuning/train_visinger2.yaml}}.


\noindent\textbf{System Annotations}
\begin{itemize}
\item \textbf{B:} The baseline model.

\item \textbf{B+D:} We add noise differentiable augmentation from scratch. The augmentation region is randomly selected as 10\% of the total interval along the temporal dimension.

\item \textbf{B\&U:} We first train the baseline for 20 epochs, then resume training the whole baseline model with an additional frame-to-sample uncertainty predictor. A uncertainty prediction loss is added using a weight of 10.0, working on the forward generator pass.

\item \textbf{B\&U\&(U+D):} We follow $\textbf{B\&U}$ for the first 80 epochs, then add \textbf{D} to the end of training.

\end{itemize}

\subsection{Corpora}
\label{sec:experiment_corpora}

We follow the official split and resample all utterances to 44.1 kHz. The corpora are introduced as follows:

\noindent \textbf{Opencpop}\cite{wang2022opencpop}. It is a Mandarin singing corpus performed by one female professional singer. It consists of 100 popular songs with corresponding MIDI annotations. All songs are recorded at 44.1kHz and have a total duration of 5.2 hours.

\noindent \textbf{Ofuton-P}\cite{ofutonp}. It is a Japanese singing corpus performed by one male singer, containing 46 songs with a total duration of 61 minutes. Most of the songs are children's rhymes. The original recording is in 96kHz, with some of them resampled to 44.1kHz before release.


\subsection{Evaluation Metrics}

We employ both objective metrics and subjective ratings to evaluate the performance of each system through VERSA toolkit~\cite{shi2025versa,shi2024versaversatileevaluationtoolkit}. We briefly introduce each metric as follows.

\textbf{Log\_F0\_RMSE}~\cite{hayashi2020espnet} measures the root mean squared error between the logarithmic fundamental frequency (F0) of the synthesized and reference singing signals, focusing on pitch accuracy. \textbf{MCD}~\cite{kubichek1993mel} quantifies the spectral distance between the generated and ground-truth audio using mel-cepstral coefficients. \textbf{Semitone} accuracy measures pitch differences between the generated sample and ground truth. \textbf{VUV} evaluates whether the model correctly predicts the voiced or unvoiced status of each frame. \textbf{Sheet-SSQA}~\cite{huang2024mos} is a pseudo MOS predictor based on a 5-point mean opinion score (MOS) scale. \textbf{$\text{MOS}_{L}$} evaluates the intelligibility and correctness of the synthesized lyrics, as rated by human listeners. \textbf{$\text{MOS}_{M}$} measures the perceived quality of melody reproduction. \textbf{$\text{MOS}$} assesses the overall naturalness of the synthesized singing voice perceived by human listeners.

For the subjective evaluation, we use MOS across different dimensions. We randomly select 30 utterances from the test set and have 20 recruited labelers rate the same set of utterances generated by different systems on a scale from 1 to 5. All labelers are native Mandarin speakers, ensuring their ability to accurately assess the quality of the synthesized lyrics. 

\begin{table*}[t]
\centering
\caption{Ablation of different types of differentiable augmentation on the Opencpop corpus in 200 epochs. B indicates baseline, and D represents differentiable augmentation. + means simultaneously use. MOS values are presented with 95\% confidence intervals.}
\vspace{-5pt}
\footnotesize
\setlength{\tabcolsep}{2.1 mm}
\begin{tabular}{lccccccccc}
\toprule
\multirow{2}{*}{Strategies} & \multicolumn{9}{c}{Opencpop} \\ \cmidrule(r){2-10} & Log\_F0\_RMSE$\downarrow$ & MCD$\downarrow$ & Semitone$\uparrow$ & VUV$\downarrow$ & Sheet-SSQA$\uparrow$ & $\text{MOS}_{L}\uparrow$ & $\text{MOS}_{M}\uparrow$ & $\text{MOS}\uparrow$ & \\ \midrule \midrule
$\textbf{B}$   & 0.174 & 7.876 & 62.78\% & 7.79\% & 4.30 & 3.65 ±0.07& 3.37 ±0.07 & 3.46 ±0.06\\
$\textbf{B+D}_{\text{Mask}}$   & \textcolor{blue}{0.161} & \textcolor{blue}{7.693} & \textcolor{blue}{63.26\%} & 7.53\% & 4.30 & \textbf{3.73 ±0.07} & 3.38 ±0.07 & \textcolor{blue}{3.49 ±0.06}\\
$\textbf{B+D}_{\text{Noise}}$  & \textbf{0.159} & \textbf{7.670} & \textbf{63.64\%} & \textcolor{blue}{7.32\%} & \textcolor{blue}{4.33} & \textcolor{blue}{3.73 ±0.07} & \textbf{3.46 ±0.07} & \textbf{3.53 ±0.06} \\
$\textbf{B+D}_{\text{Noise+Mask}}$   & 0.164 & 7.834 & 63.16\% & \textbf{7.09\%} & \textbf{4.35} & 3.70 ±0.06 & \textcolor{blue}{3.38 ±0.06} & 3.48 ±0.06\\
\midrule
\bottomrule
\end{tabular}
\label{tab_abla_diffaug}
\vspace{-2mm}
\end{table*}

\begin{table*}[t]
\centering
\caption{Ablation of different training strategies on the Ofuton-P corpus in 200 epochs. B indicates baseline, U indicates uncertainty prediction, and D represents differentiable augmentation. \& means resume and + means simultaneously use.}

\vspace{-5pt}
\footnotesize
\setlength{\tabcolsep}{2.8 mm}
\begin{tabular}{lcccccc}
\toprule
\multirow{2}{*}{Strategies} & \multicolumn{6}{c}{Ofuton-P} \\ \cmidrule(r){2-6} & Log\_F0\_RMSE$\downarrow$ & MCD$\downarrow$ & Semitone$\uparrow$ & VUV$\downarrow$ & Sheet-SSQA$\uparrow$ & \\ \midrule \midrule
$\textbf{B}$   & 0.083 & \textbf{5.690} & 66.55\% & 2.91\% & 4.19 \\
$\textbf{B+D}$   & \textcolor{blue}{0.082} & \textcolor{blue}{5.699} & 66.60\% & \textbf{2.45\%} & \textbf{4.24} \\
$\textbf{B\&U}$  & 0.082 & 5.734 & \textcolor{blue}{66.81\%} & 2.84\% & 4.21 &  \\
$\textbf{B\&U\&(U+D})$  & \textbf{0.078} & 5.751 & \textbf{66.92\%} & \textcolor{blue}{2.83\%} & \textcolor{blue}{4.22} &\\
\midrule
\bottomrule
\end{tabular}
\label{tab_ablationOfutonP}
\vspace{-2mm}
\end{table*}

\begin{figure}[t]
    \centering
\includegraphics[width=\linewidth]{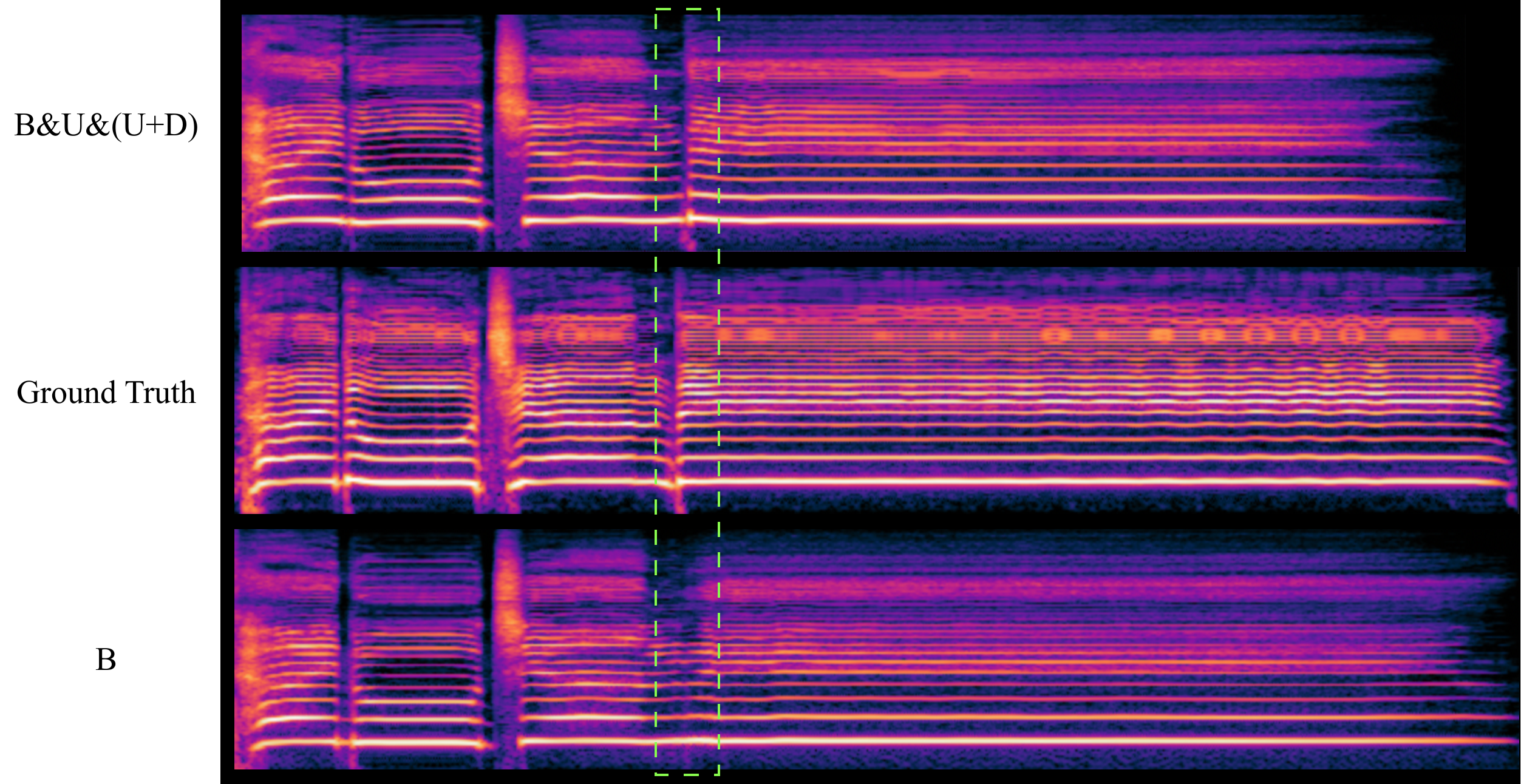}
\caption{Spectrogram case study. The green box shows the system using the uncertainty predictor, and the differentiable augmentation produces a clearer phoneme boundary, which indicates faithful lyrics.}
\label{fig_spectrogramcasestudy}
\vspace{-5mm}
\end{figure}

\subsection{Quantitative Analysis}
\label{sec:quantitativea}

We label the best performance in each metric in Table~\ref{tab_compare} using \textbf{bold font}, while marking the second place using \textcolor{blue}{blue font}. Comparing \textbf{B} and \textbf{B+D}, we observe a notable improvement in pitch and timbre-related metrics, suggesting that introducing greater variation in short intervals helps the model capture finer melodic details, which mitigates the issues mentioned in Sec.~\ref{sec:relatedwork_svs}. Noising part of the interval encourages the model to focus on other subtle artifacts that influence the perceived realism of the prior and posterior sample pairs. This enables the model to generate more expressive pitch contours and richer timbral characteristics, improving data uncertainty and obtaining a better model training process, which proves the hypothesis in Sec.~\ref{sec_diffaug}.

Adding the uncertainty predictor, as shown by results in \textbf{B\&U}, improves the perception quality in terms of pseudo MOS and MOS, in both the lyrics and melody, also overall naturalness. This shows formulating posterior uncertainty as the gap between ground truth samples and the synthesized samples, and adding a prediction module helps in efficient training. Minimizing this performance gap as an additional loss enables the model to better understand which part it mainly struggles with and improve its performance accordingly.

Collectively using differentiable augmentation and uncertainty prediction is shown to be optimal. Since this strategy has the maximum number of best or best+second-best scores across the 8 evaluation metrics and 4 systems. Notably, it improves the baseline performance in (1) \textbf{pitch}, shown by Log\_F0\_RMSE, (2) \textbf{timbre}, shown by MCD, (3) \textbf{duration}, shown by VUV, and (4) \textbf{human perception}, shown by MOS, demonstrated in Tab.~\ref{tab_compare}, mitigating the problems mentioned in Sec.~\ref{sec:relatedwork_svs}. The MOS improvement is especially significant given that the baseline is already the SOTA SVS method and does well in common cases, which further validates the effectiveness of our strategies in long-tail cases.

\subsection{Qualitative Analysis}
We show a spectrum example of the same utterance synthesized by the \textbf{B} and our best-performing system \textbf{B\&U\&(U+D)} in Fig.~\ref{fig_spectrogramcasestudy}. Zooming in on the region highlighted by the green dashed box, we can observe that the upper spectrum with differentiable augmentation and uncertainty prediction better constructs the high-energy boundary, which is a start point of a phoneme. Without a faithful modeling of this stylized region, the baseline utterance sounds ambiguous and has worse lyrics delivery. We include more examples on the website.

\section{Ablation Study}
\label{sec_ablation}

\subsection{Differentiable Augmentation Strategies}

In our main experiment, we adopt additive noise as a differentiable augmentation method. For ablation studies, we explore three strategies: additive noise, masking, and their combination, as introduced in Sec.~\ref{sec_diffaug}. In all cases, augmentation is applied to 10\% of the input features. Masking is performed either along the time or frequency dimension, depending on the discriminator type, with masked values set to zero. As shown in Table~\ref{tab_abla_diffaug}, all three strategies lead to performance improvements. Among them, using noise alone yields the best results (achieving the highest number of best or second-best scores). As noise introduces a more general form of perturbation in the feature space, encouraging the model to ignore the perturbed regions and instead focus on learning more subtle and discriminative patterns.


\subsection{Generalization to Other Languages}

We further conduct experiments on the Japanese singing corpus Ofuton-P, using differentiable augmentation and an uncertainty predictor. As described in Sec.~\ref{sec:experiment_corpora}, this corpus is significantly smaller, approximately 20\% the size of Opencpop. It mainly consists of traditional Japanese children’s songs, which are generally simple, feature repetitive phonemes, and exhibit a narrow pitch range. We train the model for 200 epochs.

The results, shown in Table~\ref{tab_ablationOfutonP}, validate the generalizability of our proposed strategies to a new corpus in another language. It demonstrates consistent improvements in Log\_F0\_MSE and Semitone, indicating enhanced pitch accuracy in the synthesized outputs. A slight degradation in MCD is observed, likely due to a trade-off between pitch precision and timbre fidelity. In addition, we see improvements in duration prediction and pseudo-MOS, reflected by VUV and Sheet-SSQA, suggesting enhanced temporal alignment and increased alignment with human perceptual preferences.

\section{Conclusion}
In this paper, we propose two uncertainty-driven strategies to enhance the training process of end-to-end singing voice synthesis (SVS) models. We incorporate prior uncertainty through differentiable data augmentation and introduce a posterior uncertainty predictor to improve robustness in long-tail scenarios. Extensive experiments demonstrate the effectiveness and generalizability of our proposed methods. In the future we will explore their collective effect with other augmentations.

\section{Acknowledgement}
Experiments of this work used the Bridges2 at PSC and Delta/DeltaAI NCSA computing systems through allocation CIS210014 from the ACCESS program, supported by NSF grants 2138259, 2138286, 2138307, 2137603, and 2138296.


{
\small
\bibliographystyle{IEEEtran}
\bibliography{custom}
}

\end{document}